\documentclass[%
reprint, 
amsmath,amssymb,
aps, physrev,
]{revtex4-2}

\usepackage{graphicx}
\usepackage{dcolumn}
\usepackage{bm}
\usepackage[colorlinks=true]{hyperref}

\usepackage{orcidlink}
\usepackage{academicons}
\newcommand{\orcid}[1]{\href{https://orcid.org/#1}{\textcolor[HTML]{A6CE39}{\orcidicon}}}

\usepackage{comment}
\usepackage{amsmath, amssymb, mathrsfs, slashed}
\usepackage{multirow, IEEEtrantools}

\usepackage{xcolor}

\begin{document}

\title{Baryon Number Susceptibilities of Hot and Dense Nuclear Matter in Relativistic Mean-Field Models}

\author{Deepak Kumar\orcidlink{0000-0001-9292-3598}}
\email{dpqraja02@gmail.com}
\affiliation{CFisUC, Department of Physics, University of Coimbra, PT 3004-516 Coimbra, Portugal}
\affiliation{Department of Physical Sciences, Indian Institute of Science Education and Research Berhampur, Berhampur, Odisha 760003, India}

\author{Biswarup Das\orcidlink{0009-0000-8486-8795}}
\email{biswarupd21@iiserbpr.ac.in}
\affiliation{Department of Physical Sciences, Indian Institute of Science Education and Research Berhampur, Berhampur, Odisha 760003, India}

\author{Sandeep Chatterjee\orcidlink{0000-0003-3914-782X}}
\email{sandeep@iiserbpr.ac.in}
\affiliation{Department of Physical Sciences, Indian Institute of Science Education and Research Berhampur, Berhampur, Odisha 760003, India}

\author{Ujjal Kumar Dey\orcidlink{0000-0002-9620-7561}}
\email{ujjal@iiserbpr.ac.in}
\affiliation{Department of Physical Sciences, Indian Institute of Science Education and Research Berhampur, Berhampur, Odisha 760003, India}

\date{\today}

\begin{abstract}
We develop a fully analytical framework for calculating baryon number susceptibilities in relativistic mean-field (RMF) models of hot and dense nuclear matter at finite temperature and baryon chemical potential. The formalism is applied to representative nonlinear RMF parameterization sets to investigate the model dependence of baryon number fluctuations over a wide range of temperatures and baryon chemical potentials. All the considered RMF parameterizations reproduce the overall qualitative beam-energy dependence of the cumulant ratios. The observed skewness and kurtosis within the RMF models satisfy the experimental values.  Specifically, the FSU-family parameterizations exhibit relatively
small deviations from the measured heavy-ion observables i.e. cumulant ratios. 
\end{abstract}


\maketitle

\section{introduction}
One of the central goals of high-energy theoretical and experimental physics is to explore the phase structure of strongly interacting matter, with particular interest in the quantum chromodynamics (QCD) critical point (CP), where the first-order phase-transition line is expected to terminate at the crossover region in the $T$--$\mu_{\rm B}$ plane~\cite{Ejiri:2007ga, Fodor:2001pe, Wilczek:1999ym, Gavai:2008zr, Datta:2016ukp}. Lattice QCD provides a first-principle framework for studying strongly interacting matter allowing for a systematic approach to investigate the QCD thermodynamics, phase structure and infer on the location of a possible CP~\cite{Ejiri:2007ga, Fodor:2001pe, Wilczek:1999ym, Gavai:2008zr, Datta:2016ukp}. 
In the vicinity of a CP, baryon number susceptibilities are expected to exhibit characteristic critical behavior, including enhanced fluctuations and non-Gaussianity. Nevertheless, the existence and precise location of the QCD CP remain unsettled. Heavy-ion collision experiments, particularly the Relativistic Heavy Ion Collider (RHIC) Beam Energy Scan (BES) program~\cite{STAR:2025zdq, STAR:2010mib, STAR:2013gus, STAR:2020tga}, provide an important experimental avenue for probing fluctuations associated with the QCD phase structure and for exploring strongly interacting matter at finite baryon density and chemical potential~\cite{Gavai:2010zn, Stephanov:2008qz, Bass:1998ca, Bhalerao:2014owa}.

Although lattice QCD provides a first-principle description of equilibrium QCD thermodynamics, its applicability at finite baryon density is restricted by the fermion sign problem, which makes simulations increasingly challenging with increasing $\mu_{\rm B}/T$~\cite{Aarts:2015tyj, Guenther:2020jwe, Borsanyi:2020fev}. This limitation motivates the use of QCD-like models, such as 
Polyakov Nambu--Jona-Lasinio (PNJL)~\cite{Liu:2026pro, Liu:2025cpj, Li:2018ygx, Bhattacharyya:2016jsn, Fukushima:2008wg, Ghosh:2006qh, Ghosh:2007wy}, Polyakov Quark Meson (PQM)~\cite{Schaefer:2007pw,Herbst:2010rf,Schaefer:2008hk, Chatterjee:2011jd} and the Walecka-type models~\cite{Walecka:1974qa, Serot:1984ey, Serot:1997xg}. Specifically, at high baryon density regime as in the core neutron stars (NS), the Walecka-type models~\cite{Walecka:1974qa, Serot:1984ey, Serot:1997xg} provide a successful framework for phenomenology. The mean field treatment of the Walecka type models particularly describe the high density nuclear matter equation of state (EOS) and thermodynamic properties of matter in neutron star interiors~\cite{Weissenborn:2011kb, Kumar:2025dlc, Kumar:2026tqe, Dutra:2014qga, Oertel:2014qza, Sanson:2026lky, kumar_2025_18013480}. These are commonly known as Relativistic Mean Field models (RMF). Calculations of baryon-number fluctuations within the RMF framework can therefore complement lattice-QCD results and provide access to the high-density regime relevant to the RHIC BES and future FAIR and NICA programs~\cite{STAR:2025zdq, CBM:2016kpk, Golovatyuk:2019rkb}.

The RMF model describes nuclear interactions through meson exchanges, with scalar ($\sigma$) and vector ($\omega$) mesons providing attractive and repulsive interactions, respectively, while the isovector ($\rho$) meson governs the isospin dependence of nuclear matter~\cite{Walecka:1974qa, Serot:1984ey, Serot:1997xg}. The corresponding coupling parameters are constrained, using the Bayesian analysis, by nuclear saturation properties, such as the binding energy per nucleon, symmetry energy, slope parameter, and incompressibility, as well as finite-nucleus observables including binding energies at nuclear saturation density ($\sim0.16$ fm$^{-3}$) and charge radii of finite nuclei~\cite{Kumar:2023rut, Kumar:2026tqe, Leonhardt:2019fua, Zhu:2022ibs, Malik:2022zol}. At higher densities, the model is extrapolated to a few times nuclear saturation density and tested against neutron-star observations, including NICER and gravitational-wave measurements~\cite{Miller:2021qha, Riley:2019yda, Miller:2019cac, LIGOScientific:2018cki}. 

In the RHIC BES program, one of the most prominent observables for probing the thermodynamic state of strongly interacting matter is event-by-event fluctuations of conserved charges, particularly the baryon number~\cite{Das:2021rck, Schmidt:2021pey}. Higher-order cumulants are sensitive to changes in the underlying thermodynamic state and may exhibit characteristic behavior near phase boundaries and critical regions~\cite{Gavai:2010zn, Stephanov:2008qz, Bass:1998ca}. Recent measurements by the STAR collaboration have reported a non-trivial higher-order baryon-number fluctuations at lower collision energies~\cite{STAR:2025zdq, STAR:2010mib, STAR:2013gus, STAR:2020tga}. Understanding the origin of this behavior requires connecting experimentally accessible fluctuation observables to the underlying interactions of strongly interacting matter. Baryon-number susceptibilities provide a natural framework for making this connection, as they quantify the response of the pressure to variations in the baryon chemical potential and encode information about the thermodynamic properties of the system~\cite{Das:2021rck, Schmidt:2021pey, Gavai:2010zn, Karsch:2010ck}. The interplay of the scalar, vector, and isovector couplings in the RMF model determines EOS and baryon number susceptibilities, making their accurate determination essential for reliable predictions of baryon-number fluctuations.

Previous studies of baryon-number fluctuations have been carried out within the hadron resonance gas~\cite{Chatterjee:2017yhp, Vovchenko:2016rkn, Alba:2014eba, Garg:2013ata, Karsch:2010ck} and RMF frameworks~\cite{Fukushima:2014lfa}. In Ref.~\cite{Fukushima:2014lfa}, author has emphasized the role of vector mediated repulsive interaction which results in the suppression in the kurtosis at small $\sqrt{s_{\rm NN}}$.
A systematic comparison of higher-order susceptibilities across a broad set of nonlinear RMF parameterizations, spanning different EOS stiffnesses and symmetry-energy behaviors, together with the latest STAR measurements, remains desirable. 
Building on Ref.~\cite{Fukushima:2014lfa}, we investigate the role of nonlinear nuclear interactions in determining higher-order baryon-number fluctuations. 
In particular, it is important to determine whether the observed suppression of $\kappa\sigma^2$ can be reproduced within conventional hadronic descriptions and how the resulting fluctuation observables depend on the underlying nuclear interaction.

In the present work, we calculate the baryon-number susceptibilities within nonlinear RMF models in thermal and chemical equilibrium which are directly related to the cumulants of the event-by-event baryon-number distribution~\cite{Das:2021rck}. 
These higher-order ratios are particularly sensitive to non-Gaussian fluctuations and to the density dependence of the underlying nuclear interaction. 
Within RMF models, the susceptibilities are determined self-consistently by the scalar and vector mean fields and therefore depend on the effective nucleon mass, vector mean fields, and nonlinear meson interactions~\cite{Walecka:1974qa, Kumar:2023rut, Menezes:2002cq, Serot:1979cc, Santos:2004js, Kumar:2026tqe}. This provides a direct connection between higher-order fluctuation observables and the microscopic structure of the effective nuclear interaction.

The remainder of the manuscript is organized as follows. In Sec.~\ref{sec:formalism}, we briefly introduce the RMF framework. The definitions and evaluation of baryon-number fluctuations are presented in Sec.~\ref{sec:fluctuations}, while the calculation of the higher-order derivatives required for the susceptibilities is discussed in Subsec.~\ref{sec:the.higher.order.derivatives}. The results and their discussion are presented in Sec.~\ref{sec:results.and.discussions}, followed by a summary and conclusions in Sec.~\ref{sec:summary.and.conclution}.

\section{Formalism} \label{sec:formalism}
Here we are focusing on the NS matter, i.e. nuclear matter at high densities described by RMF models where the hadrons are moving in the mesonic mean field background. They are interacting by exchanging of mesons. The exchange of scalar meson gives rise to the attractive interaction, while the exchange of vector meson gives rise to the repulsive interaction. We are also considering the isovector-vector meson, which defines the breaking of isospin symmetry in the system. The Lagrangian, which we are considering here, is given as follows \cite{Mishra:2001py, Tolos:2017lgv}
\begin{IEEEeqnarray}{rCl}
{\cal L} = {\cal L}_{\rm B}^{\rm kin} + {\cal L}_{\rm M}^{\rm kin} + {\cal L}_{\rm BM} - V_{\rm NL}
\label{lagrangian}
\end{IEEEeqnarray}
where, ${\cal L}_B$ is the kinetic term for the baryons given as
\begin{IEEEeqnarray}{rCl}
{\cal L}_{\rm B}^{\rm kin} = \sum_{\rm b \in {\rm B}}\bar\psi_{\rm b}(i\gamma^\mu\partial_\mu - m_{\rm b})\psi_{\rm b}. \label{lb}
\end{IEEEeqnarray}
Here, $\psi_{\rm b}$ and $m_{\rm b}$ correspond to the baryonic field and its bare mass, respectively. Similarly, the kinetic term for the mesons is given by,
\begin{IEEEeqnarray}{rCl}
{\cal L}_{\rm M}^{\rm kin} &=& \frac{1}{2} \left[ \partial_{\mu}\sigma\partial^{\mu}\sigma - m_{\sigma}^2\sigma^2 \right] - \frac{1}{4}\Omega_{\mu\nu}\Omega^{\mu\nu} + \frac{1}{2} m_{\omega}^2\omega^2 \nonumber \\
&& - \frac{1}{4}{\rm R}_{\mu\nu}{\rm R}^{\mu\nu} + \frac{1}{2}m_{\rho}^2\rho_{\mu}\rho^{\mu} 
\nonumber \\
&& \label{lmkin}
\end{IEEEeqnarray}
with $\Omega^{\mu\nu} = \partial^{\mu}\omega^{\nu} - \partial^{\nu}\omega^{\mu}$, ${\rm R}^{\mu\nu} = \partial^{\mu}{\rm \bf \rho}^{\nu} - \partial^{\nu}{\rm \bf \rho}^{\mu}$ are the mesonic field strength tensors. The $\sigma$, $\omega$ and $\rho$ meson fields are denoted by $\sigma$, $\omega$ and $\rho$ and their masses are $m_{\sigma}$, $m_{\omega}$ and $m_{\rho}$, respectively. The Lagrangian ${\cal L}_{\rm BM}$ is the Lagrangian describing the baryon meson interactions having the form,
\begin{IEEEeqnarray}{rCl}
{\cal L}_{\rm BM} &=& \sum_{{\rm b}\ \in\ {\rm B}} [\bar\psi_{\rm b}g_{\sigma {\rm b}}\sigma\psi_{\rm b} - \bar\psi_{\rm b}\gamma_{\mu}\left(g_{\omega {\rm b}}\omega^{\mu} + g_{\rho {\rm b}}{\bf \tau}_{\rm b}\cdot {\bf \rho}^{\mu} \right)\psi_{\rm b}], \nonumber \\
&&
\end{IEEEeqnarray}
where, ${\bf \tau}_{\rm b}$ is the Pauli matrix, and $g_{\alpha {\rm b}}$ for $\alpha \in \sigma, \omega^{\mu}, {\bf \rho}$ are the coupling constants of the baryons with the mesons. Similarly, $V_{\rm NL}$ describes the nonlinear interaction of mesons which is given by,
\begin{IEEEeqnarray}{rCl}
V_{\rm NL} &=& \frac{\kappa}{3!}(g_{\sigma {\rm N}}\sigma)^3 + \frac{\lambda}{4!}(g_{\sigma {\rm N}}\sigma)^4 - \frac{\xi_\omega}{4!}(g_{\omega {\rm N}}^2\omega_{\mu}\omega^{\mu})^2 \nonumber \\
&& - {\Lambda_{\omega\rho}}(g_{\omega {\rm N}}^2\omega_{\mu}\omega^{\mu})(g_{\rho {\rm N}}^2\rho_{\mu}\rho^{\mu}), \label{vnl}
\end{IEEEeqnarray}
In RMF approximation, one replaces the meson fields by their expectation values which then act as classical fields in which baryons move $i.e.$ $\langle\sigma\rangle=\sigma_0$, $\langle \omega_\mu\rangle=\omega_0\delta_{\mu 0}$, $\langle \rho_\mu^a\rangle$ =$\delta_{\mu 0}\delta_{3}^a \rho_{3}^0$. The mesonic field equations of motion can be found by the Euler-Lagrange equations for the meson fields using the Lagrangian Eq. (\ref{lagrangian})
\begin{subequations}
\begin{IEEEeqnarray}{rCl}
m_\sigma^2\sigma_0 &&+ \frac{\kappa}{2}g_{\sigma {\rm b}}^3\sigma_0^2 + \frac{\lambda}{6}g_{\sigma {\rm b}}^4\sigma_0^3 = \sum_{b} g_{\sigma {\rm b}} \rho_{{\rm s}\rm b} \label{sigma.eq} \\
m_\omega^2\omega_0 &&+ \frac{\zeta}{6}g_{\omega {\rm b}}^4\omega_0^3 + 2\Lambda_\omega (g_{\rho {\rm b}}g_{\omega {\rm b}})^2 (\rho_3^0)^2\omega_0 = \sum_{\rm b}g_{\omega {\rm b}} \rho_{{\rm v}\rm b} \nonumber \\
&& \label{omega.eq} \\
m_\rho^2\rho_3^0 &&+ 2\Lambda_\omega (g_{\rho {\rm b}}g_{\omega {\rm b}})^2 \omega_0^2\rho_3^0 = \sum_{\rm b} g_{\rho {\rm b}} I_{3\rm b}\rho_{{\rm v}\rm b}, \label{rho.eq}
\end{IEEEeqnarray}
\end{subequations}
where $\rho_{{\rm s}\rm b}$ and $\rho_{{\rm v}\rm b}$ are the scalar and vector densities for a baryon and $I_{3{\rm b}}$ is the third component of the isospin of a given baryon. We have taken $I_{3 (n,p)} = \left(-\frac{1}{2}, \frac{1}{2}\right)$. The baryon density, ($\rho_{{\rm v}\rm b}$), and scalar density, ($\rho_{{\rm s}\rm b}$), at zero temperature are given as,
\begin{IEEEeqnarray}{rCl}
\rho_{\rm vb} &=& \gamma_{\rm b} \int \frac{{\rm d}^3k}{(2\pi)^3} \left( f^{-}_{\rm b} - f^{+}_{\rm b} \right)
\\
\rho_{\rm sb} &=& \gamma_{\rm b} \int \frac{{\rm d}^3k}{(2\pi)^3} \frac{M_{\rm b}}{E_{\rm b}^{*}(k)} \left( f^{-}_{\rm b} + f^{+}_{\rm b} \right), \label{baryon.scalar.density}
\end{IEEEeqnarray}
where $E_{\rm b}^{*}(k) \equiv E_{\rm b}^{*} = \sqrt{M_{\rm b}{^2}+ k^2}$ is the single particle energy for nucleons with a medium-dependent mass given as 
\begin{equation}
M_{\rm b} = m_{\rm b} -g_{\sigma {\rm b}}\sigma_0. \label{effective.mass}
\end{equation}
The effective baryonic chemical potential, $\tilde{\mu}_{\rm b}$, is given as,
\begin{equation}
\tilde{\mu}_{\rm b} = {\mu}_{\rm b} - g_{\omega {\rm b}}\omega_0 - g_{\rho {\rm b}}I_{\rm 3b}\rho_3^0. \label{effective.chemical.potential}
\end{equation}
Further, $\gamma_{\rm b}=2$ corresponds to the spin degeneracy factor for nucleons. The energy density, $\epsilon$, within the RMF model is given by
\begin{IEEEeqnarray}{rCl}
\epsilon &=& \gamma_{\rm b} \sum_{\rm b} \int \frac{d^3k}{(2\pi)^3} E_{\rm b}^{*} \left(f_{\rm b}^{-} + f_{\rm b}^{+}\right) \\
&& + \frac{1}{2}m_{\sigma}^2\sigma_0^2 + \frac{1}{2} m_{\omega}^2\omega_0^2 + \frac{1}{2} m_{\rho}^2{\rho_{3}^0}^2 + \frac{\kappa}{3!}(g_{\sigma{\rm N}}\sigma_0)^3 \nonumber
\\
&& + \frac{\lambda}{4!}(g_{\sigma}\sigma_0)^4 + \frac{\xi_{\omega}}{4!}(g_{\omega}\omega)^4 + 3{\Lambda_{\omega\rho}}(g_{\rho}g_{\omega}\rho_0 \omega_0)^2
\nonumber
\\
&& \label{total.energy.density.nm}
\end{IEEEeqnarray}

\noindent Similarly, the pressure, $P$, is obtained as 
\begin{IEEEeqnarray}{rCl}
P &=& \sum_{\rm b} P_{\rm b} - \frac{1}{2}m_\sigma^2\sigma_0^2 + \frac{1}{2}m_\omega^2\omega_0^2 + \frac{1}{2}m_\rho^2(\rho_3^0)^2 + P_{\rm int}, \nonumber \\
&& \label{total.pressure.nm}
\end{IEEEeqnarray}
where the baryon contribution to the pressure ($P_{\rm b}$) and the pressure due to interaction term ($P_{\rm int}$) are given as following 
\begin{IEEEeqnarray}{rCl}
P_{\rm b} &=& \frac{\gamma_{\rm b}}{\beta}\int \frac{d^3k}{(2\pi)^3}
\left[
\ln\left(1+e^{-\beta\omega_{\rm b}^{+}}\right) + \ln\left(1+e^{-\beta\omega_{\rm b}^{-}}\right)
\right], \nonumber\\
&&
\label{baryon.pressure}
\end{IEEEeqnarray}
\begin{IEEEeqnarray}{rCl}
P_{\rm int} &=& - \frac{\kappa}{6}(g_{\sigma{\rm b}}\sigma_0)^3 - \frac{\lambda}{24}(g_{\sigma{\rm b}}\sigma_0)^4 + \frac{\zeta}{24}(g_{\omega{\rm b}}\omega_0)^4 \nonumber \\
&& + \Lambda_\omega(g_{\rho{\rm b}}g_{\omega{\rm b}})^2 (\rho_3^0)^2\omega_0^2, \label{interaction.pressure}
\end{IEEEeqnarray}
where $\beta=1/T$ and $\omega_{\rm b}^{\pm} = E_{\rm b}^{*} \mp \tilde{\mu}_{\rm b}$ and the fermi-dirac distribution functions $f_{\rm b}^{\pm}$ are defined as
\begin{equation}
f_{\rm b}^{\pm}= \frac{1}{1 + e^{\beta \omega_{\rm b}^{\pm}}}.
\end{equation}
The NS matter is charge neutral as well as $\beta$-equilibrated, i.e. $n \to p + e$ and $n \to p + \mu$, where $n$, $p$, $e$ and $\mu$ denote neutron, proton, electron and muon, respectively. We have taken the leptonic contribution to get the EOS for the core of NS at zero temperature limit. 
On the other hand the chemical potential of baryons are defined as
\begin{IEEEeqnarray}{rCl}
{\mu}_b = b_b{\mu}_B + q_b {\mu}_Q + s_b {\mu}_S, \label{chemical.potential}
\end{IEEEeqnarray}
where, $\mu_B$, $\mu_Q$ and $\mu_S$ are the baryonic, electric and strange chemical potentials and $b_b$, $q_b$ and $s_b$ are the baryon, charge and strange quantum numbers, respectively. In the present work, in the context of RHIC, we consider ${\mu}_Q = 0$ and ${\mu}_S = 0$. By this choice, the value of the iso-vector mean field comes to zero.

\section{Baryon number susceptibility in RMF model}
\label{sec:fluctuations}
We follow the standard convention as used in Ref. \cite{Fukushima:2014lfa} for thermal fluctuations, which are derived from the derivatives of the pressure with respect to the relevant chemical potentials and here in the present study, the baryon chemical potential. For the baryon number fluctuation, thus, we calculate the dimensionless baryon number susceptibilities defined as
\begin{equation}
\chi_{\rm B}^{(n)} = T^{\,n-4} \frac{\partial^n {P}}{\partial\mu_{\rm B}^n},
\qquad n=1,2,3, \dots.
\label{chi.definition}
\end{equation}
Here, $P$ denotes the thermodynamic pressure, while $\mu_{\rm B}$ and $T$ are the baryon chemical potential and temperature, respectively. 
Since the temperature is held fixed while differentiating with respect to the baryon chemical potential, Eq.~(\ref{chi.definition}) reduces to
\begin{IEEEeqnarray}{rCl}
\chi_{\rm B}^{(1)} &=& \frac{P^{(1)}}{T^3}, \quad \chi_{\rm B}^{(2)} = \frac{P^{(2)}}{T^2}, \nonumber \\
\chi_{\rm B}^{(3)} &=& \frac{P^{(3)}}{T},   \quad \chi_{\rm B}^{(4)} = P^{(4)}.
\label{chi.derivatives}
\end{IEEEeqnarray}
where, we denote $P^{(n)} = \frac{\partial^n {P}}{\partial\mu_{\rm B}^n}$. The first two cumulants correspond to the mean baryon number $M$ and variance $\sigma^2$, while the third and fourth cumulants are related to the 
skewness ($S\sigma$) and kurtosis ($K\sigma^2$), 
\begin{IEEEeqnarray}{rCl}
S\sigma = \frac{{\chi}_B^{(3)}}{{\chi}_B^{(2)}}, \label{skewness} \qquad
K\sigma^2 = \frac{{\chi}_B^{(4)}}{{\chi}_B^{(3)}}  \label{Kurtosis}
\end{IEEEeqnarray}

We obtain the $n{\rm th}$ derivatives of pressure $P^{(n)}$ for $n=1,\dots,4$ using Eq.~(\ref{total.pressure.nm}), which require the relevant derivatives of meson fields and their $n{\rm th}$ derivatives i.e. $\sigma_0^{(i)}$, $\omega_0^{(i)}$, and ${\rho_3^0}^{(i)}$ for $i=0,n$ where $i=0$ denotes meson field values without taking derivative as given in Eqs. (\ref{sigma.eq})~-~(\ref{rho.eq}) while $i=n$ denotes the $n{\rm th}$ derivative of the same with respect to baryon chemical potential. The higher order derivatives of meson fields ($\sigma_0^{(n)}$, $\omega_0^{(n)}$, and ${\rho_3^0}^{(n)}$) are obtained by solving the equations obtained from  Eqs. (\ref{sigma.eq})~-~(\ref{rho.eq}) by taking respective derivatives with respect to baryon chemical potential. In the following section, we discuss these higher order derivatives in detail. 

\subsection{The higher order derivatives of pressure and meson fields} \label{sec:the.higher.order.derivatives} 
Here, we give a detailed calculation of finding the higher order susceptibilities within the RMF model of nuclear matter at high density and temperature. In this model, baryons are taken as fermions and the interactions between baryons are governed by exchange of mesons. Scalar meson causes attractive interaction while vector meson gives repulsive interaction. We consider a generalized Lagrangian as given in Eq. (\ref{lagrangian}), where $\psi_b$ denotes baryons while $\sigma$ ($\omega$) denotes scalar (vector) meson field. The symbol $\rho$ in the Lagrangian denotes isovector meson field. We also consider other various interaction terms. In the mean field approximation, the mean field values of mesons are found by minimizing the thermodynamic potential $\Omega = -P$ which gives the equations of motion for meson fields as Eqs. (\ref{sigma.eq})~-~(\ref{rho.eq}). The expression for pressure is given in Eqs. (\ref{total.pressure.nm})~-~(\ref{interaction.pressure}) and the expressions of the dimensionless baryon number susceptibilities are given in Eq. (\ref{chi.definition}) and Eq. (\ref{chi.derivatives}). The derivatives of the total pressure with respect to $\mu_{B}$ which are included in the higher order susceptibilities are given as
\begin{IEEEeqnarray}{rCl}
P^{(n)} &=& \sum_{b} P_b^{(n)} - \frac{m_\sigma^2}{2}(\sigma_0^2)^{(n)} + \frac{m_\omega^2}{2}(\omega_0^2)^{(n)} \nonumber \\
&& + \frac{m_\rho^2}{2}\left[(\rho_3^0)^2\right]^{(n)} + P_{\rm int}^{(n)}, \quad n=1,\dots,4, \label{pressure.derivatives}
\end{IEEEeqnarray}
with
\begin{IEEEeqnarray}{rCl}
P_{\rm int}^{(n)} &=& - \frac{\kappa}{6}g_{\sigma{\rm b}}^3(\sigma_0^3)^{(n)} - \frac{\lambda}{24}g_{\sigma{\rm b}}^4 (\sigma_0^4)^{(n)} + \frac{\zeta}{24}g_{\omega{\rm b}}^4 (\omega_0^4)^{(n)} \nonumber \\
&& + \Lambda_\omega (g_{\rho{\rm b}}g_{\omega{\rm b}})^2 \left[(\rho_3^0)^2\omega_0^2\right]^{(n)}. \label{interaction.pressure.derivatives}
\end{IEEEeqnarray}

The kinetic contribution to the baryonic pressure is given in Eq. (\ref{baryon.pressure}). Its derivatives with respect to the baryon chemical potential are
\begin{IEEEeqnarray}{rCl}
P_{\rm b}^{(1)} &=& -\frac{1}{\pi^2} \int_0^\infty k^2
\left[
f_{\rm b}\left(\omega_{\rm b}^{+}\right)^{(1)} + \bar f_{\rm b}\left(\omega_{\rm b}^{-}\right)^{(1)}
\right] dk, \nonumber \\
&&
\label{pressure1}
\end{IEEEeqnarray}
\begin{IEEEeqnarray}{rCl}
P_{\rm b}^{(2)} &=& \frac{1}{\pi^2} \int_0^\infty k^2
\Bigg[
\beta f_{\rm b}(1-f_{\rm b}) 
\left[\left(\omega_{\rm b}^{+}\right)^{(1)}\right]^2 - f_{\rm b}\left(\omega_{\rm b}^{+}\right)^{(2)} \nonumber\\
&&
+ \beta\bar f_{\rm b}(1-\bar f_{\rm b})
\left[\left(\omega_{\rm b}^{-}\right)^{(1)}\right]^2 - \bar f_{\rm b}\left(\omega_{\rm b}^{-}\right)^{(2)}
\Bigg] dk, \label{pressure2}
\end{IEEEeqnarray}
\begin{IEEEeqnarray}{rCl}
P_{\rm b}^{(3)} &=& \frac{1}{\pi^2} \int_0^\infty k^2
\Bigg[
-\beta^2f_{\rm b}(1-f_{\rm b})(1-2f_{\rm b})
\left[\left(\omega_{\rm b}^{+}\right)^{(1)}\right]^3
\nonumber\\
&&
+ 3\beta f_{\rm b}(1-f_{\rm b}) \left(\omega_{\rm b}^{+}\right)^{(1)} \left(\omega_{\rm b}^{+}\right)^{(2)}
- f_{\rm b}\left(\omega_{\rm b}^{+}\right)^{(3)} \nonumber\\
&&
-\beta^2\bar f_{\rm b}(1-\bar f_{\rm b})(1-2\bar f_{\rm b}) \left[\left(\omega_{\rm b}^{-}\right)^{(1)}\right]^3 \nonumber\\
&&
+ 3\beta\bar f_{\rm b}(1-\bar f_{\rm b}) \left(\omega_{\rm b}^{-}\right)^{(1)} \left(\omega_{\rm b}^{-}\right)^{(2)}
- \bar f_{\rm b}\left(\omega_{\rm b}^{-}\right)^{(3)}
\Bigg] dk, \nonumber \\
&& \label{pressure3}
\end{IEEEeqnarray}
\begin{IEEEeqnarray}{rCl}
P_{\rm b}^{(4)} &=& \frac{1}{\pi^2} \int_0^\infty k^2
\Bigg[
\beta^3f_{\rm b}(1-f_{\rm b}) (1-6f_{\rm b}+6f_{\rm b}^2) \left[\left(\omega_{\rm b}^{+}\right)^{(1)}\right]^4 \nonumber\\
&&
- 6\beta^2f_{\rm b}(1-f_{\rm b})(1-2f_{\rm b}) \left[\left(\omega_{\rm b}^{+}\right)^{(1)}\right]^2 \left(\omega_{\rm b}^{+}\right)^{(2)} \nonumber\\
&&
+ 3\beta f_{\rm b}(1-f_{\rm b}) \left[\left(\omega_{\rm b}^{+}\right)^{(2)}\right]^2 \nonumber \\
&&
+ 4\beta f_{\rm b}(1-f_{\rm b}) \left(\omega_{\rm b}^{+}\right)^{(1)} \left(\omega_{\rm b}^{+}\right)^{(3)}
- f_{\rm b}\left(\omega_{\rm b}^{+}\right)^{(4)} 
\nonumber\\
&&
+ \beta^3\bar f_{\rm b}(1-\bar f_{\rm b}) (1-6\bar f_{\rm b}+6\bar f_{\rm b}^2) \left[\left(\omega_{\rm b}^{-}\right)^{(1)}\right]^4 \nonumber\\
&&
- 6\beta^2\bar f_{\rm b}(1-\bar f_{\rm b})(1-2\bar f_{\rm b}) \left[\left(\omega_{\rm b}^{-}\right)^{(1)}\right]^2 \left(\omega_{\rm b}^{-}\right)^{(2)} \nonumber\\
&&
+ 3\beta\bar f_{\rm b}(1-\bar f_{\rm b}) \left[\left(\omega_{\rm b}^{-}\right)^{(2)}\right]^2 \nonumber \\
&&
+ 4\beta\bar f_{\rm b}(1-\bar f_{\rm b}) \left(\omega_{\rm b}^{-}\right)^{(1)} \left(\omega_{\rm b}^{-}\right)^{(3)}
- \bar f_{\rm b}\left(\omega_{\rm b}^{-}\right)^{(4)}
\Bigg] dk. \nonumber \\
&&
\label{pressure4}
\end{IEEEeqnarray}
where $\left(\omega_{\rm b}^{\pm}\right)^{(n)} = E_{\rm b}^{*(n)} \mp \tilde{\mu}_{\rm b}^{(n)}$ for $n=1,\dots,4$. The Eqs.~(\ref{pressure1})~-~(\ref{pressure4}), contain the derivatives of meson fields. The derivatives of the quadratic, cubic and quartic self-interaction terms of mesons fields, $\phi = (\sigma_0, \omega_0,\, \rho_3^0)$ are given as,
\begin{subequations}
\begin{IEEEeqnarray}{rCl}
(\phi^2)^{(1)}&=&2\phi\phi^{(1)}, \\
(\phi^2)^{(2)}&=&2\left[(\phi^{(1)})^2+\phi\phi^{(2)}\right], \\
(\phi^2)^{(3)}&=&2\left[3\phi^{(1)}\phi^{(2)}+\phi\phi^{(3)}\right], \\
(\phi^2)^{(4)}&=&2\left[3(\phi^{(2)})^2+4\phi^{(1)}\phi^{(3)}+\phi\phi^{(4)}\right], \label{phi2.derivatives}
\end{IEEEeqnarray}
\end{subequations}
\begin{subequations}
\begin{IEEEeqnarray}{rCl}
(\phi^3)^{(1)} &=& 3\phi^2\phi^{(1)}, \\
(\phi^3)^{(2)} &=& 6\phi(\phi^{(1)})^2 + 3\phi^2\phi^{(2)}, \\
(\phi^3)^{(3)} &=& 6(\phi^{(1)})^3 + 18\phi\phi^{(1)}\phi^{(2)} + 3\phi^2\phi^{(3)}, \\
(\phi^3)^{(4)} &=& 36(\phi^{(1)})^2\phi^{(2)} + 24\phi\phi^{(1)}\phi^{(3)} + 18\phi(\phi^{(2)})^2 \nonumber \\ 
&& + 3\phi^2\phi^{(4)}, \label{phi3.derivatives}
\end{IEEEeqnarray}
\end{subequations}
and
\begin{subequations}
\begin{IEEEeqnarray}{rCl}
(\phi^4)^{(1)} &=& 4\phi^3\phi^{(1)}, \\
(\phi^4)^{(2)} &=& 12\phi^2(\phi^{(1)})^2 + 4\phi^3\phi^{(2)}, \\
(\phi^4)^{(3)} &=& 24\phi(\phi^{(1)})^3 + 36\phi^2\phi^{(1)}\phi^{(2)} + 4\phi^3\phi^{(3)}, \\
(\phi^4)^{(4)} &=& 24(\phi^{(1)})^4 + 144\phi(\phi^{(1)})^2\phi^{(2)} \\
&& + 36\phi^2(\phi^{(2)})^2 + 48\phi^2\phi^{(1)}\phi^{(3)} + 4\phi^3\phi^{(4)}. \nonumber \\
&& \label{phi4.derivatives}
\end{IEEEeqnarray}
\end{subequations}
and the mixed interaction term derivative can be evaluated as 
\begin{equation}
\left[(\rho_3^0)^2\omega_0^2\right]^{(n)} = (AB)^{(n)} = \sum_{k=0}^{n} \binom{n}{k} A^{(k)}B^{(n-k)}, \label{AB.derivatives}
\end{equation}
where $n=1,\dots,4$ corresponding to first, second, third and fourth derivatives. Now these field derivatives can be obtained by solving the meson field equations as given in Eq. (\ref{sigma.eq}), (\ref{omega.eq}), and (\ref{rho.eq}) and the equations for their derivatives with respect to $\mu_{B}$ as defined below,
\begin{subequations}
\begin{IEEEeqnarray}{rCl}
m_\sigma^2\sigma_0^{(n)} &&+ 
\kappa g_{\sigma{\rm b}}^3 \left(\frac{\sigma_0^2}{2}\right)^{(n)} + \frac{\lambda}{2}g_{\sigma{\rm b}}^4 \left(\frac{\sigma_0^3}{3}\right)^{(n)} = \sum_{b} g_{\sigma{\rm b}} \rho_{{\rm s}\rm b}^{(n)}, \nonumber \\
&& \label{sigma.eq.n}
\\
m_\omega^2\omega_0^{(n)} &&+
\frac{\zeta}{2}g_{\omega{\rm b}}^4 \left(\frac{\omega_0^3}{3}\right)^{(n)} \nonumber \\
+&& 2\Lambda_\omega (g_{\rho{\rm b}}g_{\omega{\rm b}})^2 \left[(\rho_3^0)^2\omega_0\right]^{(n)} = \sum_{\rm b} g_{\omega{\rm b}} \rho_{{\rm v}\rm b}^{(n)}, \label{omega.eq.n}
\\
m_\rho^2(\rho_3^0)^{(n)} &&+
2\Lambda_\omega (g_{\rho{\rm b}}g_{\omega{\rm b}})^2 \left[\omega_0^2\rho_3^0\right]^{(n)} = \sum_{\rm b} g_{\rho{\rm b}} I_{3b} \rho_{{\rm v}\rm b}^{(n)}. \label{rho.eq.n}
\end{IEEEeqnarray}
\end{subequations}

Now the derivatives of scalar densities and number densities are given as following. The scalar density of baryons can also be written as
\begin{equation}
\rho_{{\rm s}\rm b} = \frac{1}{\pi^2} \int_0^\infty k^2 A_{\rm b} \left(f_{\rm b}+\bar f_{\rm b}\right) \,dk. \label{scalar.density}
\end{equation}
Here, we define $A_{\rm b}\equiv \frac{M_{\rm b}}{E_{\rm b}^*}$ for simplicity and further we give its different derivatives with respect to baryon chemical potential. The derivatives of the scalar density with respect to the baryon chemical potential are as following
\begin{subequations}
\begin{IEEEeqnarray}{rCl}
\rho_{{\rm s}\rm b}^{(1)} &=& \frac{1}{\pi^2} \int_0^\infty k^2
\left[
A_{\rm b}^{(1)} \left(f_{\rm b}+\bar f_{\rm b}\right)
+
A_{\rm b}\left(f_{\rm b}^{(1)}+\bar f_{\rm b}^{(1)}\right)
\right] dk, \nonumber \\
&& 
\end{IEEEeqnarray}
\begin{IEEEeqnarray}{rCl}
\rho_{{\rm s}\rm b}^{(2)} &=& \frac{1}{\pi^2} \int_0^\infty k^2
\left[
A_{\rm b}^{(2)} \left(f_{\rm b}+\bar f_{\rm b}\right)
+
2A_{\rm b}^{(1)} \left(f_{\rm b}^{(1)}+\bar f_{\rm b}^{(1)}\right)\right. \nonumber \\
&& \left.+ A_{\rm b} \left(f_{\rm b}^{(2)}+\bar f_{\rm b}^{(2)}\right)
\right] dk,
\end{IEEEeqnarray}
\begin{IEEEeqnarray}{rCl}
\rho_{{\rm s}\rm b}^{(3)} &=& \frac{1}{\pi^2} \int_0^\infty k^2
\left[
A_{\rm b}^{(3)} \left(f_{\rm b}+\bar f_{\rm b}\right)
+
3A_{\rm b}^{(2)} \left(f_{\rm b}^{(1)}+\bar f_{\rm b}^{(1)}\right)
\right.
\nonumber\\
&&
\left. +
3A_{\rm b}^{(1)} \left(f_{\rm b}^{(2)}+\bar f_{\rm b}^{(2)}\right)
+
A_{\rm b} \left(f_{\rm b}^{(3)}+\bar f_{\rm b}^{(3)}\right) 
\right] dk, \nonumber \\
&&
\end{IEEEeqnarray}
\begin{IEEEeqnarray}{rCl}
\rho_{{\rm s}\rm b}^{(4)} &=& \frac{1}{\pi^2} \int_0^\infty k^2
\left[
A_{\rm b}^{(4)} \left(f_{\rm b}+\bar f_{\rm b}\right)
+ 
4A_{\rm b}^{(3)} \left(f_{\rm b}^{(1)}+\bar f_{\rm b}^{(1)}\right)
\right.
\nonumber\\
&&
+ 6A_{\rm b}^{(2)} \left(f_{\rm b}^{(2)}+\bar f_{\rm b}^{(2)}\right)
+ 4A_{\rm b}^{(1)} \left(f_{\rm b}^{(3)}+\bar f_{\rm b}^{(3)}\right) \nonumber \\
&&
\left. + A_{\rm b} \left(f_{\rm b}^{(4)}+\bar f_{\rm b}^{(4)}\right) \right] dk,
\label{scalar.density.derivatives}
\end{IEEEeqnarray}
\end{subequations}
where $f_{\rm b}^{(n)}$ and $\bar f_{\rm b}^{(n)}$ are the $n^{\rm th}$ derivative of fermi distributions of particles and anti-particles with respective to the baryon chemical potential. We give their expression in Eqs. (\ref{fd1})~-~(\ref{fd4}). Similar to the derivatives of scalar density, we can obtain the derivatives of number density of baryons. The number density is given by
\begin{equation}
\rho_{{\rm v}\rm b} = \frac{1}{\pi^2} \int_0^\infty k^2 \left(f_{\rm b}-\bar f_{\rm b}\right) \,dk. \label{vector.density}
\end{equation}
Its derivatives with respect to the baryon chemical potential are
\begin{IEEEeqnarray}{rCl}
\rho_{{\rm v}\rm b}^{(n)} &=& \frac{1}{\pi^2} \int_0^\infty k^2 \left( f_{\rm b}^{(n)} - \bar f_{\rm b}^{(n)} \right) \,dk, \label{vector.density.derivatives}
\end{IEEEeqnarray}

The particle and antiparticle Fermi-Dirac distribution functions are as follows
\begin{equation}
f_{\rm b} = \frac{1}{e^{\beta\omega_{\rm b}^{+}}+1},
\qquad
\bar f_{\rm b} = \frac{1}{e^{\beta\omega_{\rm b}^{-}}+1}, \label{fermi.dist}
\end{equation}

The derivatives of the distribution functions with respect to the baryon chemical potential are
\begin{subequations}
\begin{IEEEeqnarray}{rCl}
f_{\rm b}^{(1)} &=& 
-\beta f_{\rm b}(1-f_{\rm b}) \left(\omega_{\rm b}^{+}\right)^{(1)}, \label{fd1}
\end{IEEEeqnarray}
\begin{IEEEeqnarray}{rCl}
f_{\rm b}^{(2)} &=& 
-\beta f_{\rm b}(1-f_{\rm b}) \left(\omega_{\rm b}^{+}\right)^{(2)} \nonumber \\
&& + \beta^2f_{\rm b}(1-f_{\rm b})(1-2f_{\rm b}) \left[\left(\omega_{\rm b}^{+}\right)^{(1)}\right]^2,  \label{fd2}
\end{IEEEeqnarray}
\begin{IEEEeqnarray}{rCl}
f_{\rm b}^{(3)} &=&
-\beta^3f_{\rm b}(1-f_{\rm b})(1-6f_{\rm b}+6f_{\rm b}^2)
\left[\left(\omega_{\rm b}^{+}\right)^{(1)}\right]^3
\nonumber\\
&&
+ 3\beta^2f_{\rm b}(1-f_{\rm b})(1-2f_{\rm b})
\left(\omega_{\rm b}^{+}\right)^{(1)}
\left(\omega_{\rm b}^{+}\right)^{(2)}
\nonumber\\
&&
- \beta f_{\rm b}(1-f_{\rm b}) \left(\omega_{\rm b}^{+}\right)^{(3)},  \label{fd3}
\end{IEEEeqnarray}
\begin{IEEEeqnarray}{rCl}
f_{\rm b}^{(4)}
&=&
\beta^4f_{\rm b}(1-f_{\rm b})
(1-14f_{\rm b}+36f_{\rm b}^2-24f_{\rm b}^3)
\left[\left(\omega_{\rm b}^{+}\right)^{(1)}\right]^4
\nonumber\\
&& - 6\beta^3f_{\rm b}(1-f_{\rm b})
(1-6f_{\rm b}+6f_{\rm b}^2)
\left[\left(\omega_{\rm b}^{+}\right)^{(1)}\right]^2
\left(\omega_{\rm b}^{+}\right)^{(2)}
\nonumber\\
&&
+ 3\beta^2f_{\rm b}(1-f_{\rm b})(1-2f_{\rm b})
\left[\left(\omega_{\rm b}^{+}\right)^{(2)}\right]^2 \nonumber \\
&&
+ 4\beta^2f_{\rm b}(1-f_{\rm b})(1-2f_{\rm b})
\left(\omega_{\rm b}^{+}\right)^{(1)}
\left(\omega_{\rm b}^{+}\right)^{(3)}
\nonumber\\
&&
- \beta f_{\rm b}(1-f_{\rm b})
\left(\omega_{\rm b}^{+}\right)^{(4)}.
 \label{fd4}
\end{IEEEeqnarray}
\end{subequations}

The antiparticle derivatives $\bar f_{\rm b}$ are obtained by replacing $\omega_{\rm b}^{+}$ by $\omega_{\rm b}^{-}$. The derivatives of effective mass ($M_{\rm b}$), and effective chemical potential ($\tilde{\mu}_{\rm b}$) are required to define all these definitions. Now, we are giving the expressions for effective mass, effective chemical potential and their derivatives with respect to $\mu_{B}$. These are defined as 
\begin{IEEEeqnarray}{rCl}
M_{\rm b} &=& m_{\rm b} - g_{\sigma{\rm b}}\sigma_0, \\
M_{\rm b}^{(n)} &=& - g_{\sigma{\rm b}}\sigma_0^{(n)}, \label{effective.mass}
\end{IEEEeqnarray}
\begin{IEEEeqnarray}{rCl}
\tilde{\mu}_{\rm b} &=& \mu_{\rm b} - g_{\omega{\rm b}}\omega_0 - g_{\rho{\rm b}}I_{\rm 3b}\rho_3^0, \\
\tilde{\mu}_{\rm b}^{(n)} &=& \mu_{\rm b}^{(n)} - g_{\omega{\rm b}}\omega_0^{(n)} - g_{\rho{\rm b}}I_{\rm 3b}\left(\rho_3^0\right)^{(n)}, \label{effective.chemical.potential}
\end{IEEEeqnarray}

The single particle energy ($E_{\rm b}^*$) of a baryon is given by
\begin{equation}
E_{\rm b}^*=\sqrt{k^2+M_{\rm b}^2}. \label{effective.energy}
\end{equation}
Its derivatives with respect to the $\mu_{B}$ are
\begin{subequations}
\begin{IEEEeqnarray}{rCl}
E_{\rm b}^{*(1)} &=& \frac{M_{\rm b}}{E_{\rm b}^*}M_{\rm b}^{(1)},
\end{IEEEeqnarray}
\begin{IEEEeqnarray}{rCl}
E_{\rm b}^{*(2)} &=& 
\frac{k^2}{E_{\rm b}^{*3}}\left(M_{\rm b}^{(1)}\right)^2 + \frac{M_{\rm b}}{E_{\rm b}^*}M_{\rm b}^{(2)},
\end{IEEEeqnarray}
\begin{IEEEeqnarray}{rCl}
E_{\rm b}^{*(3)} &=& 
-3\frac{k^2M_{\rm b}}{E_{\rm b}^{*5}}\left(M_{\rm b}^{(1)}\right)^3
+3\frac{k^2}{E_{\rm b}^{*3}}M_{\rm b}^{(1)}M_{\rm b}^{(2)}
+\frac{M_{\rm b}}{E_{\rm b}^*}M_{\rm b}^{(3)}, \nonumber \\
&&
\end{IEEEeqnarray}
\begin{IEEEeqnarray}{rCl}
E_{\rm b}^{*(4)} &=& 
-3\frac{k^2(E_{\rm b}^{*2}-5M_{\rm b}^2)}{E_{\rm b}^{*7}} \left(M_{\rm b}^{(1)}\right)^4 \nonumber \\
&& - 18\frac{k^2M_{\rm b}}{E_{\rm b}^{*5}} \left(M_{\rm b}^{(1)}\right)^2M_{\rm b}^{(2)} \nonumber\\
&& + 3\frac{k^2}{E_{\rm b}^{*3}} \left(M_{\rm b}^{(2)}\right)^2 + 4\frac{k^2}{E_{\rm b}^{*3}} M_{\rm b}^{(1)}M_{\rm b}^{(3)} + \frac{M_{\rm b}}{E_{\rm b}^*}M_{\rm b}^{(4)}. \nonumber \\
&& \label{effective.energy.derivatives}
\end{IEEEeqnarray}
\end{subequations}

Now, we calculate the derivative of $A_{\rm b} = \frac{M_{\rm b}}{E_{\rm b}^*}$ in the following as
\begin{subequations}
\begin{IEEEeqnarray}{rCl}
A_{\rm b}^{(1)} &=& \frac{M_{\rm b}^{(1)}}{E_{\rm b}^*} - \frac{M_{\rm b}E_{\rm b}^{*(1)}}{E_{\rm b}^{*2}}, \end{IEEEeqnarray}
\begin{IEEEeqnarray}{rCl}
A_{\rm b}^{(2)}
&=&
\frac{M_{\rm b}^{(2)}}{E_{\rm b}^*}
-2\frac{M_{\rm b}^{(1)}E_{\rm b}^{*(1)}}{E_{\rm b}^{*2}}
+2\frac{M_{\rm b}\left(E_{\rm b}^{*(1)}\right)^2}{E_{\rm b}^{*3}}
-\frac{M_{\rm b}E_{\rm b}^{*(2)}}{E_{\rm b}^{*2}}, \nonumber \\
&&
\end{IEEEeqnarray}
\begin{IEEEeqnarray}{rCl}
A_{\rm b}^{(3)}
&=&
\frac{M_{\rm b}^{(3)}}{E_{\rm b}^*}
-3\frac{M_{\rm b}^{(2)}E_{\rm b}^{*(1)}}{E_{\rm b}^{*2}}
+6\frac{M_{\rm b}^{(1)}\left(E_{\rm b}^{*(1)}\right)^2}{E_{\rm b}^{*3}} \nonumber \\
&&
-3\frac{M_{\rm b}^{(1)}E_{\rm b}^{*(2)}}{E_{\rm b}^{*2}}
-6\frac{M_{\rm b}\left(E_{\rm b}^{*(1)}\right)^3}{E_{\rm b}^{*4}}
+6\frac{M_{\rm b}E_{\rm b}^{*(1)}E_{\rm b}^{*(2)}}{E_{\rm b}^{*3}} \nonumber\\
&&
-\frac{M_{\rm b}E_{\rm b}^{*(3)}}{E_{\rm b}^{*2}},
\end{IEEEeqnarray}

\begin{IEEEeqnarray}{rCl}
A_{\rm b}^{(4)}
&=&
\frac{M_{\rm b}^{(4)}}{E_{\rm b}^*}
-4\frac{M_{\rm b}^{(3)}E_{\rm b}^{*(1)}}{E_{\rm b}^{*2}}
+12\frac{M_{\rm b}^{(2)}\left(E_{\rm b}^{*(1)}\right)^2}{E_{\rm b}^{*3}}
\nonumber\\
&&
-6\frac{M_{\rm b}^{(2)}E_{\rm b}^{*(2)}}{E_{\rm b}^{*2}}
-24\frac{M_{\rm b}^{(1)}\left(E_{\rm b}^{*(1)}\right)^3}{E_{\rm b}^{*4}}
\nonumber\\
&&
+24\frac{M_{\rm b}^{(1)}E_{\rm b}^{*(1)}E_{\rm b}^{*(2)}}{E_{\rm b}^{*3}}
-4\frac{M_{\rm b}^{(1)}E_{\rm b}^{*(3)}}{E_{\rm b}^{*2}} 
\nonumber\\
&&
+24\frac{M_{\rm b}\left(E_{\rm b}^{*(1)}\right)^4}{E_{\rm b}^{*5}}
\nonumber\\
&&
-36\frac{M_{\rm b}\left(E_{\rm b}^{*(1)}\right)^2E_{\rm b}^{*(2)}}{E_{\rm b}^{*4}}
+6\frac{M_{\rm b}\left(E_{\rm b}^{*(2)}\right)^2}{E_{\rm b}^{*3}}
\nonumber\\
&&
+8\frac{M_{\rm b}E_{\rm b}^{*(1)}E_{\rm b}^{*(3)}}{E_{\rm b}^{*3}}
-\frac{M_{\rm b}E_{\rm b}^{*(4)}}{E_{\rm b}^{*2}}.
\label{Ab.derivatives}
\end{IEEEeqnarray}

\end{subequations}

\begin{table*}
\centering
\caption{RMF parameter sets employed in the present work. Meson masses ($m_\sigma$, $m_\omega$ and $m_\rho$) and the coupling $\kappa$ are in MeV unit while others are dimensionless.}
\begin{tabular}{lcccccccccc}
\toprule
Model & $m_\sigma$ & $m_\omega$ & $m_\rho$ & $g_\sigma$ & $g_\omega$ & $g_\rho$ & $\kappa$ & $\lambda$ & $\xi_{\omega}$ & $\Lambda_{\omega\rho}$ \\
\hline
NL3 & 508.194 & 782.501 & 763 & 10.2169 & 12.8675 & 8.948 & -0.01963 & -0.01585 & 0.0 & 0.0 \\
GM1 & 512 & 783 & 770 & 8.91 & 10.61 & 8.196 & 0.01565 & -0.000006 & 0.0 & 0.0 \\
TM1 & 511.198 & 783 & 770 & 10.0289 & 12.6139 & 9.2644 & -0.01434 & 0.000367 & 0.0169 & 0.0 \\
TM1e & 511.198 & 783 & 770 & 10.0289 & 12.6139 & 13.9714 & -0.01434 & 0.000367 & 0.0169 & 0.0429 \\
FSU2 & 497.479 & 782.5 & 763 & 10.3968 & 13.5569 & 8.9703 & 3.0029 & -0.000533 & 0.0256 & 0.000823 \\
FSU2R & 497.479 & 782.5 & 763 & 10.3718 & 13.5054 & 14.3675 & 3.0911 & -0.00168 & 0.024 & 0.045 \\
FSU & 491.5 & 782.5 & 763 & 10.5924 & 14.302 & 11.7673 & 1.4203 & 0.023762 & 0.06 & 0.03 \\
\hline\hline
\end{tabular}
\label{tab:rmf_parameters}
\end{table*}

\section{Results and discussions} \label{sec:results.and.discussions}
\begin{figure}
    \centering
    \begin{minipage}{0.45\textwidth}
    \centering
    \includegraphics[width=\textwidth]{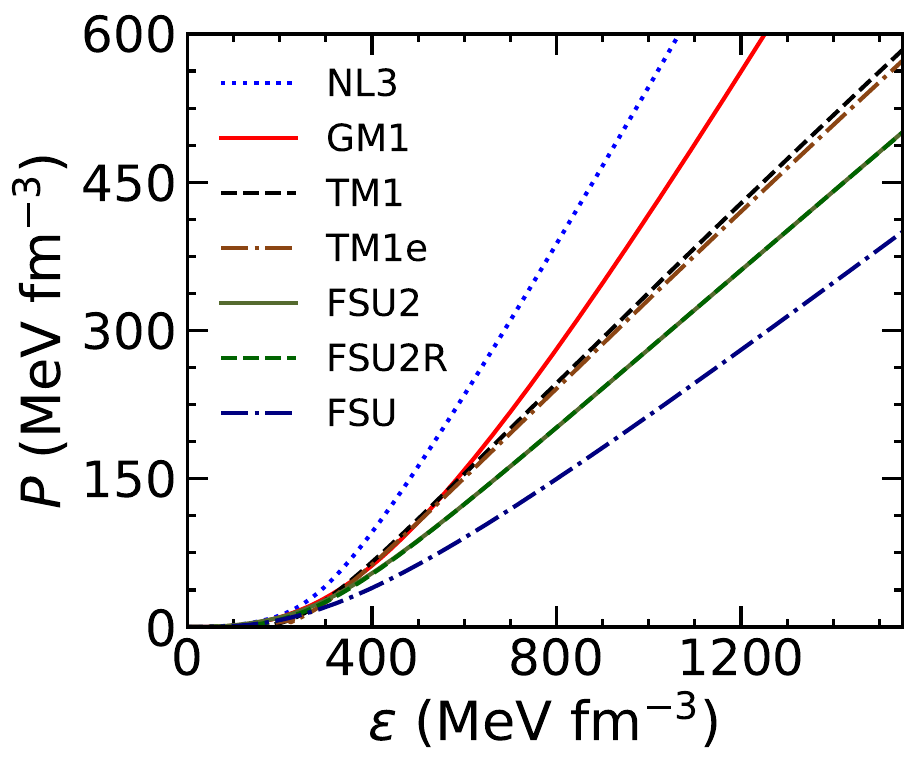}
    \centerline{(a)}
    \end{minipage}
    \hfill
    \begin{minipage}{0.43\textwidth}
    \centering
    \includegraphics[width=\textwidth]{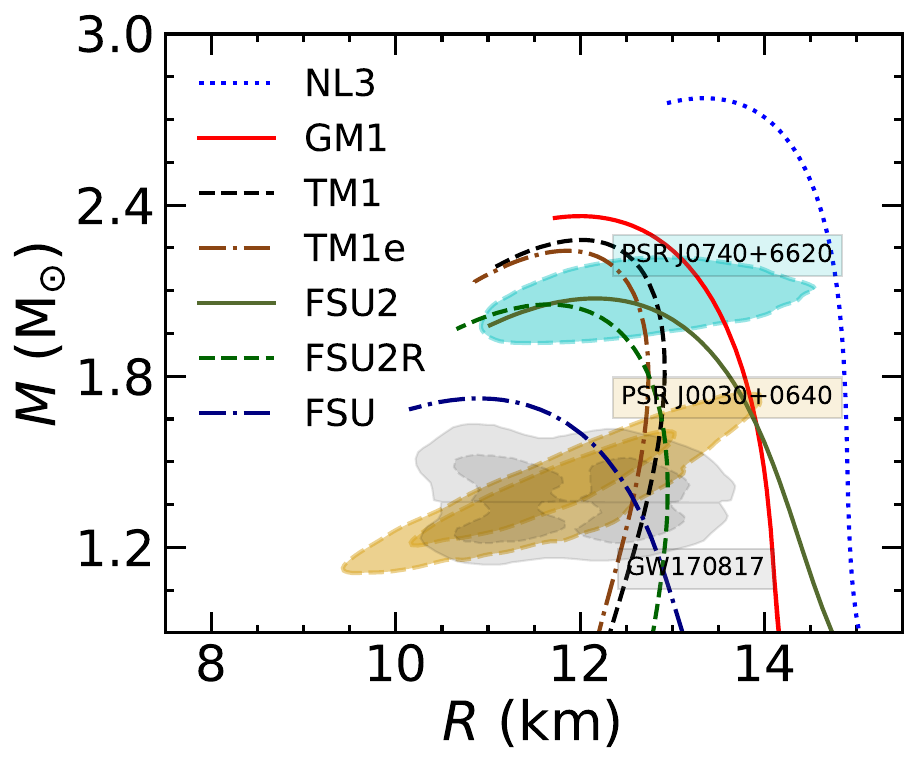}
    \centerline{(b)}
    \end{minipage}
    \caption{Panel (a): The pressure, ($P$), is shown as a function of the energy density, ($\varepsilon$), illustrating the differences in the stiffness of the EOS among the models at zero temperature. Panel (b): Tolman--Oppenheimer--Volkoff (TOV) equation solutions as mass-radius curves for corresponding EOS. The golden and teal colored shaded regions are the NICER observations of pulsars PSR J0740+6620 \cite{Dittmann:2024mbo} and PSR J0030+0640 \cite{Vinciguerra:2023qxq} respectively. The gray region is the gravitational wave observation GW170817 \cite{LIGOScientific:2017vwq, LIGOScientific:2017ync, LIGOScientific:2018cki}. 
    \label{fig:eos_and_amr}}
\end{figure}

In this section, we investigate the temperature and baryon chemical potential dependence of higher-order baryon-number fluctuations within the RMF model framework. Our primary focus is on the ratios of cumulants of baryon number fluctuations i.e. $\chi_3/\chi_1$, $\chi_3/\chi_2$ and $\chi_4/\chi_2$. We consider here a broad range of RMF model parameter sets as listed in Table~\ref{tab:rmf_parameters}, including the stiff EOS models NL3~\cite{Lalazissis:1996rd, Lalazissis:2009zz}, GM1~\cite{Glendenning:1991es, Flores:2013yqa}, the moderately stiff models TM1~\cite{Shen:1998gq, Shen:2020sec} and TM1e~\cite{Shen:2020sec}, the soft models FSU~\cite{Todd-Rutel:2005yzo, Fattoyev:2010mx}, FSU2~\cite{Chen:2014sca}, and FSU2R~\cite{Tolos:2017lgv}, which span a range of nuclear-matter properties and EOS stiffnesses. These parameter sets are calibrated to reproduce the saturation properties of nuclear matter and observational constraints on neutron stars at zero temperature, while exhibiting distinct high-density behaviors. The corresponding EOSs are shown in Fig.~\ref{fig:eos_and_amr}(a) and the mass-radius curves for NSs are shown in Fig.~\ref{fig:eos_and_amr}(b), illustrating the broad range of stiffness spanned by these models at zero temperature limits. All these models except NL3 produce the properties of nuclear matter at saturation density and follow the NS observations from gravitational waves (GW) and NICER observatories. This allows us to examine how the microscopic interaction structure encoded in different RMF parameterizations influences higher-order fluctuations. We subsequently confront the predictions from the above parameterizations resulting with the available net-proton fluctuation measurements from the RHIC BES \cite{STAR:2025zdq, STAR:2013gus}.

These models differ mainly in the values of the meson-baryon couplings and nonlinear interaction terms. The conventional stiff EOS models NL3, GM1, TM1, and TM1e lead to a stronger repulsive interactions at high densities and hence a larger NS radii. In contrast, the moderately soft models FSU, FSU2, FSU2R, which incorporate finite  $\xi_{\omega}$ ($\omega$ meson self interaction) and $\Lambda_{\omega\rho}$ ($\omega$-$\rho$ meson interaction) couplings constants as seen in Eq.~(\ref{vnl}), are more compatible with recent astrophysical constraints from NICER and GW observations as seen in Fig.~\ref{fig:eos_and_amr}(b). These models span a wide range of nuclear matter properties and provide a suitable framework for examining the sensitivity of the EOS and higher-order baryon number fluctuations to the underlying interactions.

The recent BES measurements by the STAR Collaboration provide the ratios of baryon-number cumulants over a range of collision energies $\sqrt{s_{\rm NN}}$~\cite{STAR:2014egu, STAR:2025zdq, STAR:2013gus}. The thermal conditions, $T$ and $\mu_{B}$ at freeze-out are obtained by HRG model analysis of hadron yields as below~\cite{Cleymans:2005xv, Andronic:2008gu, Andronic:2005yp}
\begin{align}
T(\mu_B) &= a + b\,\mu_B^2 + c\,\mu_B^4, \label{chemi_freeze_temp} \\
\mu_B(\sqrt{s_{\rm NN}}) &= \frac{d}{1+e\,\sqrt{s_{\rm NN}}}, \label{chemi_freeze_amub}
\end{align}
where the parameters are $a=0.166~{\rm GeV}$, $b=0.139~{\rm GeV}^{-1}$, $c=0.053~{\rm GeV}^{-3}$, $d=1.308~{\rm GeV}$, and $e=0.273~{\rm GeV}^{-1}$. For every $\sqrt{s_{\rm NN}}$, we have computed baryon number susceptibility within the RMF models at corresponding $T$ and $\mu_{B}$ as given by Eqs.~(\ref{chemi_freeze_temp}) and (\ref{chemi_freeze_amub}).

\begin{figure}
    \centering
    \begin{minipage}{0.45\textwidth}
    \centering
    \includegraphics[width=\textwidth]{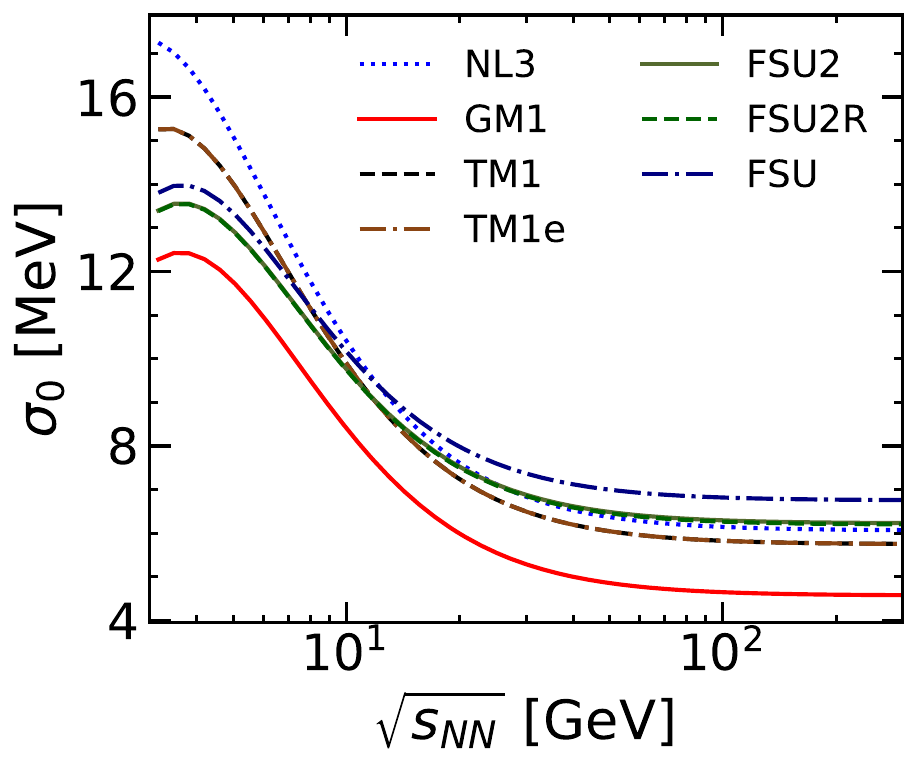}
    \centerline{(a)}
    \end{minipage}
    \hfill
    \begin{minipage}{0.45\textwidth}
    \centering
    \includegraphics[width=\textwidth]{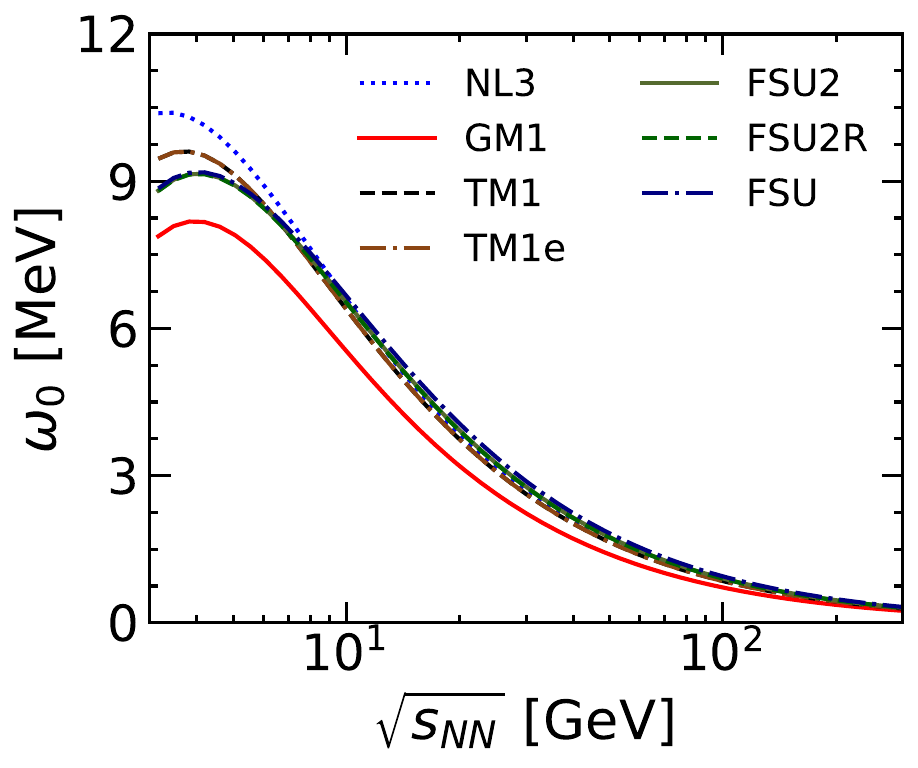}
    \centerline{(b)}
    \end{minipage}
    \caption{\label{fig:sigma_and_omega} 
    Evolution of the mean meson fields as functions of beam energy, ($\sqrt{s_{\rm NN}}$), for the RMF parameterizations considered in this work. Top Panel (a) shows a variation of the scalar ($\sigma_0$) field with ($\sqrt{s_{\rm NN}}$), reflecting the strength of the attractive scalar interaction and the corresponding reduction of the effective baryon mass in dense matter. Bottom Panel (b) displays the behavior of the vector ($\omega_0$) field as a function of ($\sqrt{s_{\rm NN}}$), characterizing the repulsive vector interaction associated with the net-baryon density. Here, the $\rho_0$ meson field is zero as we consider isospin symmetric matter.
    }
\end{figure}
In Fig.~\ref{fig:sigma_and_omega} we present the variation of the mean scalar field $\sigma_0$ and vector field $\omega_0$ as functions of the beam energy  $\sqrt{s_{\rm NN}}$ for the various RMF parameterizations considered in this work. Here it is important to mention that the mean iso-vector field $\rho_0$ vanishes as we consider isospin symmetric dense matter. Despite the differences in the underlying meson couplings and nonlinear interaction terms, all models exhibit remarkably similar behavior, leading to a substantial overlap of the corresponding curves. 
Both fields values decrease with increasing $\sqrt{s_{\rm NN}}$ and attain a peak near smaller $\sqrt{s_{\rm NN}}\sim 5\ {\rm MeV}$. As shown in the top panel, the differences among the models are more pronounced at lower values of $\sqrt{s_{\rm NN}}$ for both $\sigma_0$ and $\omega_0$. At higher collision energies, the model dependence for $\omega_0$ becomes negligible, whereas noticeable differences for $\sigma_0$ persist. These differences in the meson-field behavior are expected to influence baryon-number fluctuations and the corresponding cumulant ratios.
The rise (decline) of the scalar field reflects the increasing (decreasing) scalar attraction and the corresponding reduction (increase) of the effective nucleon mass in dense matter. Similarly, the rise (decline) of vector field increases (reduces) the repulsion in the dense matter at smaller $\sqrt{s_{\rm NN}}$. We investigate these effects in detail in the following sections.

\begin{figure*}
    \centering
    \begin{minipage}{0.32\textwidth}
    \centering
    \includegraphics[width=\textwidth]{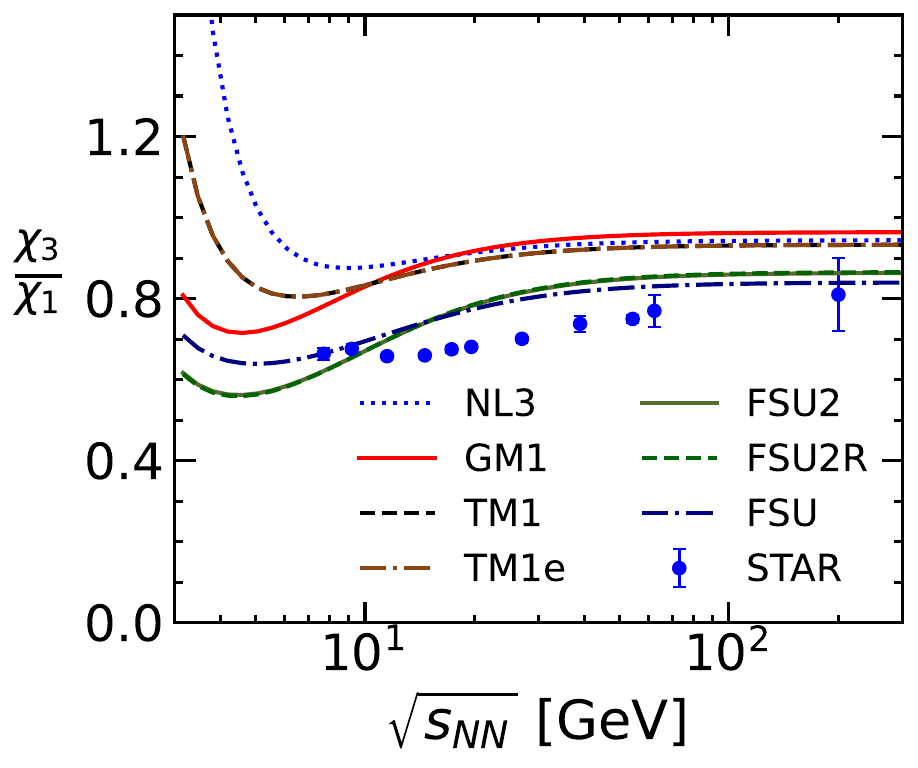}
    \centerline{(a)}
    \end{minipage}
    \hfill
    \begin{minipage}{0.32\textwidth}
    \centering
    \includegraphics[width=\textwidth]{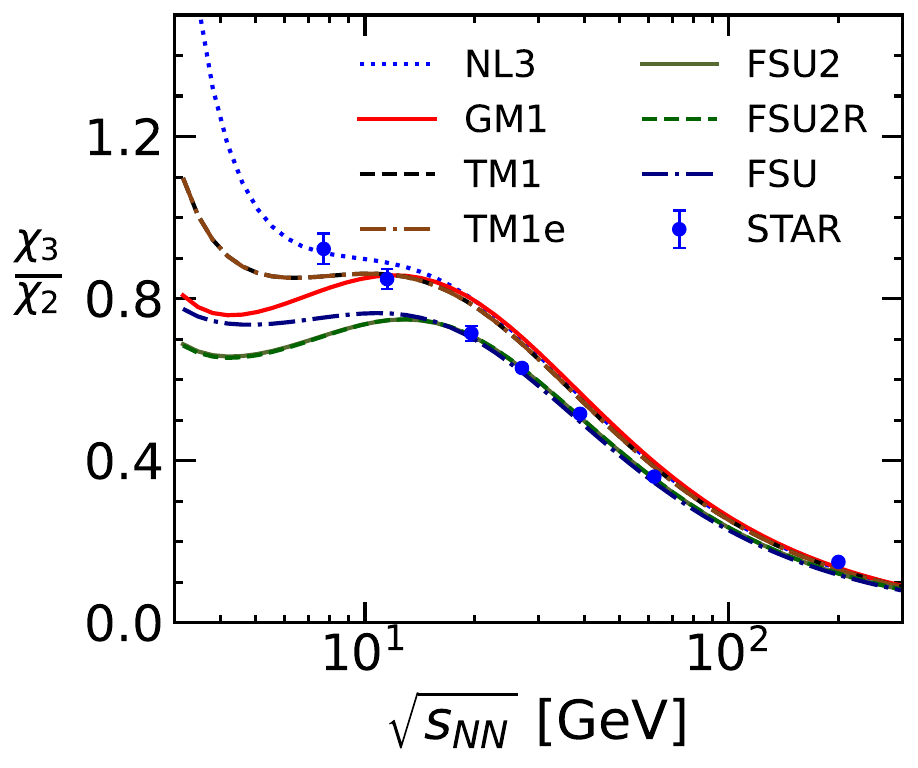}
    \centerline{(b)}
    \end{minipage}
    \hfill
    \begin{minipage}{0.32\textwidth}
    \centering
    \includegraphics[width=\textwidth]{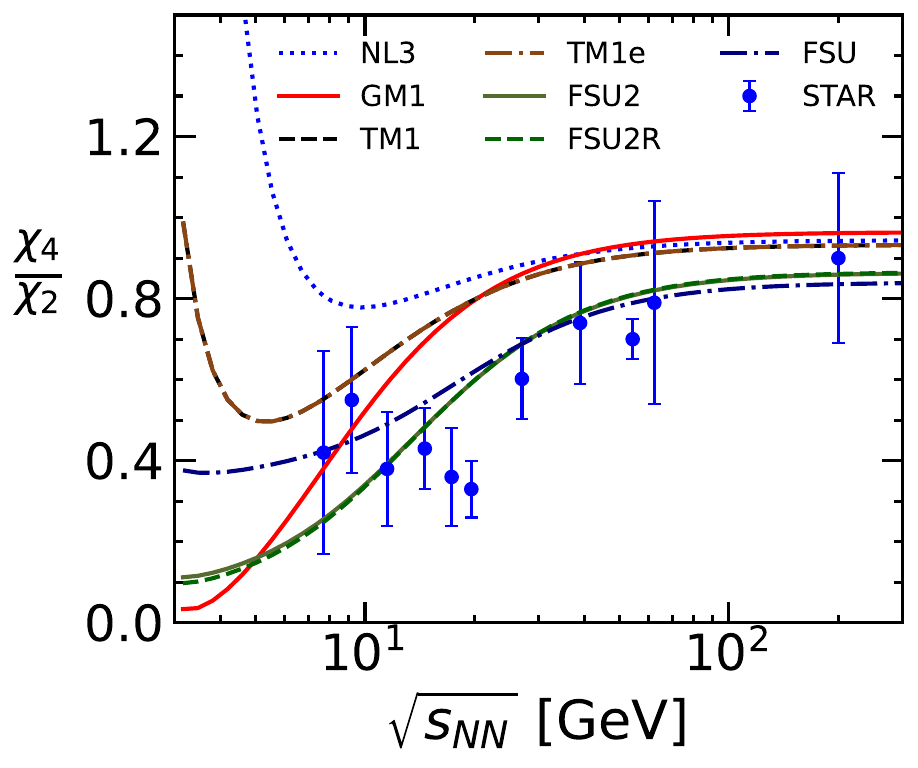}
    \centerline{(c)}
    \end{minipage}
    \caption{Collision-energy dependence of the cumulant ratios, ($\chi_3/\chi_1$, $\chi_3/\chi_2$ and $\chi_4/\chi_2$), obtained within the RMF models along the chemical freeze-out curve. Panel \ref{fig:cumulant_ratios}(a) shows the variation of the $\chi_3/\chi_1$ with beam energy, ($\sqrt{s_{\rm NN}}$). Panel \ref{fig:cumulant_ratios}(b) shows the variation of $S\sigma = \chi_3/\chi_2$ with $\sqrt{s_{\rm NN}}$ which illustrates a gradual suppression of net-baryon density fluctuations with increasing collision energy. Panel \ref{fig:cumulant_ratios}(c) presents the corresponding behavior of the kurtosis $\kappa\sigma^2 = \chi_4/\chi_2$, which exhibits a non-monotonic dependence on ($\sqrt{s_{\rm NN}}$) and approaches the Poisson limit, $\kappa\sigma^2 \simeq 1$, at high energies. Blue dots with vertical lines represent the STAR observations with errors \cite{STAR:2025zdq, STAR:2013gus}.
    \label{fig:cumulant_ratios}
    }
\end{figure*}
Figure \ref{fig:cumulant_ratios} displays the collision-energy dependence of the various cumulant ratios e.g. $\chi_3/\chi_1$, $\chi_3/\chi_2$ and $\chi_4/\chi_2$ in Figs.~\ref{fig:cumulant_ratios}(a), \ref{fig:cumulant_ratios}(b) and \ref{fig:cumulant_ratios}(c) respectively. Here, we also display the corresponding experimental measurements at a given beam energy $\sqrt{s_{\rm NN}}$~\cite{STAR:2025zdq, STAR:2013gus, STAR:2014egu}. 
The ratio $\chi_3/\chi_1$ as shown in Fig.~\ref{fig:cumulant_ratios}(a) exhibits similar beam energy dependence across all RMF parameterizations considered in this work, particularly at higher beam energies $\sqrt{s_{\rm NN}}$ i.e. high $T$ and small $\mu_{B}$. Initially, for higher value of $\sqrt{s_{\rm NN}}$ this ratio increases monotonically with increasing $\sqrt{s_{\rm NN}}$ and saturates for all the models. Conversely, at lower beam energies $\sqrt{s_{\rm NN}}$, corresponding to lower $T$ and higher $\mu_{B}$, the ratio $\chi_3/\chi_1$ decreases monotonically with increasing $\sqrt{s_{\rm NN}}$, reaching a minimum near $\sqrt{s_{\rm NN}} \sim 5$~GeV. 

This non-monotonic behavior is consistently predicted by all RMF models examined in this study, though with quantitative differences and model-dependent variations. Future measurements from the STAR fixed-target program and the NICA facility at lower collision energies could provide stringent constraints on the predicted behavior of $\chi_3/\chi_1$ and help identify the RMF parameterizations that are most consistent with the experimental data.
The skewness $S\sigma=\chi_3/\chi_2$, shown in Fig.~\ref{fig:cumulant_ratios}(b), exhibits a monotonic decrease with increasing beam energy $\sqrt{s_{\rm NN}}$, attaining relatively large values at low collision energies and gradually approaching smaller values toward higher $\sqrt{s_{\rm NN}}$. This trend is consistent with previous experimental measurements~\cite{STAR:2013gus}. From a physical perspective, the observed decrease in $S\sigma$ reflects the evolution of the system from a baryon-rich regime at low beam energies toward a progressively more baryon-symmetric medium at higher collision energies.
Next, we see the variation of kurtosis $\kappa\sigma^2 = \chi_4/\chi_2$ as a function of $\sqrt{s_{\rm NN}}$ in Fig.~\ref{fig:cumulant_ratios}(c) for different RMF model parameterizations. A pronounced non-monotonic behavior of $\kappa\sigma^2$ can be seen in the model. It starts from a large value at very low $\sqrt{s_{\rm NN}}$, and decreases to a minima in the intermediate beam energies and subsequently increases, approaching to a saturation near unity at high $\sqrt{s_{\rm NN}} \sim 200$~GeV. Such a trend indicates a smooth transition from a strongly interacting baryon-dominated medium to a regime where the fluctuations become increasingly consistent with the expectations of an ideal hadron gas \cite{Fukushima:2014lfa}. We find that our results are in qualitative agreement with the recent experimental observations reported by the STAR Collaboration~\cite{STAR:2025zdq}. {At very low collision energies, $\sqrt{s_{\rm NN}}\lesssim 5$~GeV, $\kappa\sigma^2$ exhibits pronounced model-dependent behavior, providing enhanced sensitivity to the differences among the RMF parameterizations. Further experimental measurements in this low-energy regime will be important for constraining the theoretical predictions and assessing the differences among the RMF parameterizations.}

The predictions {for the various ratios obtained} from the various RMF parameterizations are found to be similar {at high $\sqrt{s_{\rm NN}} \sim 200$~GeV}, demonstrating that the ratios are insensitive to the details of the underlying nuclear interaction at such $\sqrt{s_{\rm NN}}$ i.e. high $T$ and low $\mu_{B}$. These models differ significantly near $\sqrt{s_{\rm NN}} \sim 5$~GeV and have a discriminatory information among the RMF models considered here. The RMF calculations reproduce the qualitative behavior observed in the experimental data, namely the decreasing trend of $S\sigma$ and the non-monotonic evolution of $\kappa\sigma^2$, with the latter approaching to a saturation near one $\kappa\sigma^2 \simeq 1$, at sufficiently high collision energies. The largest deviations in the low-energy region, where the effects of strong interactions and high net-baryon densities are expected to be most pronounced. 

Using the available experimental measurements~\cite{STAR:2013gus, STAR:2014egu, STAR:2025zdq}, we quantify the agreement between the model predictions and the experimental data by evaluating the average deviation $\Delta_i$ at various $\sqrt{s_{\rm NN}}$ to compare the estimate from different RMF model parameterizations, which satisfy nuclear saturation properties, astrophysical observations of NS as well as BES/STAR observations. Here, $\Delta_i$ is defined as follows,  
\begin{equation}
\Delta_i = \frac{1}{N} \sum_j^N \frac{|\mathcal{O}_{ij}^{\rm mod} - \mathcal{O}_{ij}^{\rm exp}|}{\mathcal{O}_{ij}^{\rm exp}{}}, \label{delta}
\end{equation}
where $i$ stands for the observable and $j$ runs for the $\sqrt{s_{\rm NN}}$ values at which the observable is determined experimentally. 
The $\mathcal{O}_{ij}^{\rm mod}$ ($\mathcal{O}_{ij}^{\rm exp}$) represents the model predicted value (experimentally observed value in BES/STAR) of cumulant ratios i.e. $i = \chi_3/\chi_1,\ \chi_3/\chi_2\ {\rm and}\ \chi_4/\chi_2$ at the $j$-th collision energy and $N$ is the total number of $\sqrt{s_{\rm NN}}$ values.

\begin{table}
  \centering
  \caption{Numerical values of average $\Delta$ obtained for each RMF parameterization from comparisons with the STAR measurements of the ratios $\chi_3/\chi_1$, $\chi_3/\chi_2$ and $\chi_4/\chi_2$. The parameterization with the minimum $\Delta$ provides the best overall agreement with the experimental data.}
    \begin{tabular}{lrrrr}
    \hline\hline
    Model & \multicolumn{1}{l}{$\Delta_{\left(\frac{\chi_3}{\chi_1}\right)}$} & \multicolumn{1}{l}{$\Delta_{\left(\frac{\chi_3}{\chi_2}\right)}$} & \multicolumn{1}{l}{$\Delta_{\left(\frac{\chi_4}{\chi_2}\right)}$} & $\sum_i = \Delta_i$\\
    \hline
    NL3 & 0.283 & 0.081 & 0.641 & 1.005 \\
    GM1 & 0.268 & 0.094 & 0.424 & 0.785 \\
    TM1 & 0.246 & 0.077 & 0.466 & 0.789 \\
    TM1e & 0.246 & 0.077 & 0.466 & 0.789 \\
    FSU2 & 0.106 & 0.084 & 0.210 & 0.401 \\
    FSU2R& 0.108 & 0.083 & 0.213 & 0.404 \\
    FSU & 0.083 & 0.087 & 0.201 & 0.371 \\
    \hline\hline
    \end{tabular}%
  \label{tab:chisq_values}%
\end{table}%

\begin{figure}
    \centering
    \begin{minipage}{0.95\columnwidth}
    \centering
    \includegraphics[width=\textwidth]{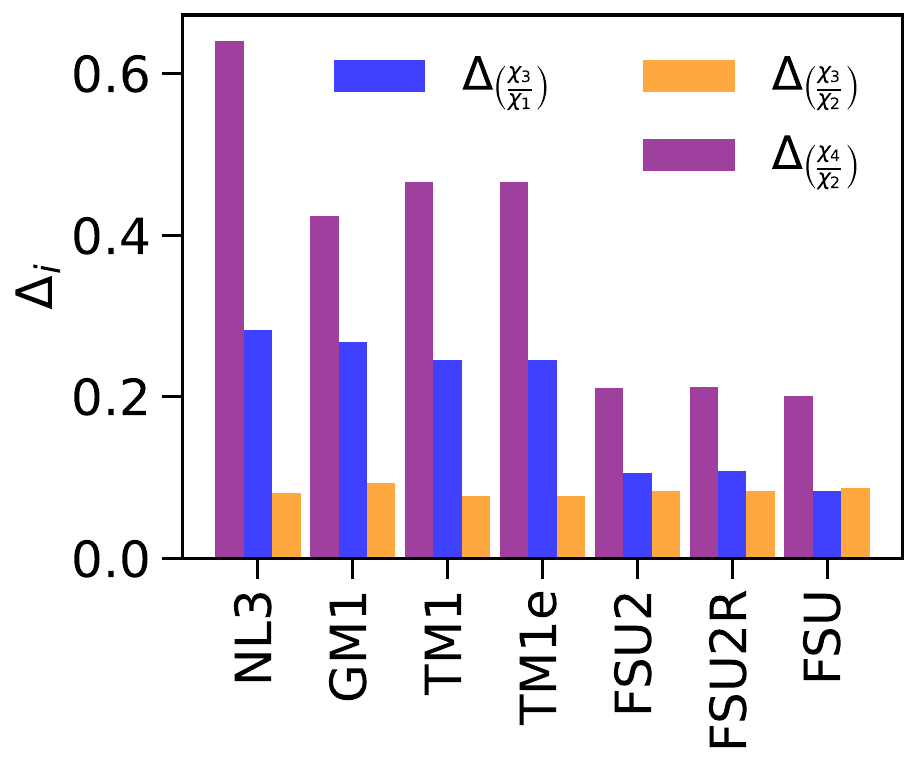}
    \end{minipage}
    \caption{Comparison of the deviation $\Delta_i$ for different RMF models obtained from the BES/STAR measurements of $i =  \chi_3/\chi_1,\ \chi_3/\chi_2\ {\rm and}\ \chi_4/\chi_2$. For each observable, the deviation $\Delta_i$ is computed as the sum of absolute differences between the model predictions and the experimental data at given $\sqrt{s_{\rm NN}}$~\cite{STAR:2013gus, STAR:2014egu, STAR:2025zdq}. Smaller value of $\Delta_i$ indicate better agreement with the measurements, allowing identification of the RMF model that best reproduces the STAR results. \label{fig:chisq_bar}
    }
\end{figure}
{Figure~\ref{fig:chisq_bar}  presents a comparative assessment of the different RMF parameterizations based on their average deviations $\Delta_i$ from the experimental measurements. The deviations are evaluated separately for the three cumulant ratios, $\chi_3/\chi_1$, $\chi_3/\chi_2$, and $\chi_4/\chi_2$, over various values of $\sqrt{s_{\rm NN}}$. The resulting average $\Delta_i$ values are presented as a bar diagram, enabling a quantitative comparison of the model predictions with the available experimental data.}

Different heights of bars represent a numerical value as given in Table~\ref{tab:chisq_values}. For each RMF model, the deviation $\Delta_i$ is evaluated by using Eq.~(\ref{delta}) i.e. the sum of absolute differences between the model predictions and the experimental data at given $\sqrt{s_{\rm NN}}$ over all available beam energies, $\sqrt{s_{\rm NN}}$~\cite{STAR:2013gus, STAR:2014egu, STAR:2025zdq}. {The bars in Fig.~\ref{fig:chisq_bar} corresponding to cumulant ratios provide a quantitative measure of the agreement between RMF models and the BES/STAR measurements~\cite{STAR:2013gus, STAR:2014egu, STAR:2025zdq}.} Smaller value of $\Delta_i$ indicates a better agreement with the experimental data, thereby enabling a direct comparison of the predictive performance of the RMF models. Among the RMF parameterizations considered, the FSU family exhibits relatively smaller average deviations $\Delta_i$ from the available RHIC measurements, indicating a closer description of the experimentally observed cumulant ratios. Furthermore, in the astrophysical point of view, within this family, the FSU2R parameterization provides a small average $\Delta_i$ and satisfies the NS observational constraints as discussed in Fig.~\ref{fig:eos_and_amr} and nuclear-matter saturation properties.

\section{Summary and conclusion} \label{sec:summary.and.conclution}
In this work, we have investigated the higher-order cumulant ratios of net-baryon number fluctuations, namely $\chi_3/\chi_1$, $\chi_3/\chi_2$, and $\chi_4/\chi_2$, within a relativistic mean-field (RMF) description of dense baryonic matter composed of neutrons and protons. We have derived the analytical expressions for the higher-order baryon-number susceptibilities within the RMF framework, allowing the response of the thermodynamic system to changes in the baryon chemical potential to be studied consistently up to fourth order. These higher-order susceptibilities provide information on the non-Gaussian characteristics of baryon-number fluctuations and are therefore particularly useful for examining the density dependence of the nuclear interaction over the range of thermodynamic conditions relevant to heavy-ion collisions.

To explore the sensitivity of these observables to the underlying nuclear interaction, we have considered a large set of RMF parameterizations, namely NL3, GM1, TM1, TM1e, FSU, FSU2, and FSU2R. These parameterizations span a broad range of nuclear-matter incompressibility, symmetry-energy behavior, and high-density EOS stiffness, thereby allowing us to examine the model dependence of the fluctuation observables. Our results show that all the considered RMF parameterizations reproduce the overall qualitative beam-energy dependence of the cumulant ratios. However, substantial quantitative differences emerge among the models, particularly in the lower beam-energy regime where the system probes larger baryon chemical potentials and higher baryon densities. These differences originate from the different behavior of the density dependence of scalar and vector meson fields in the various parameterizations. The stiffer interactions, such as NL3 and GM1, generally exhibit stronger mean-field effects and consequently larger deviations in the predicted cumulant ratios, whereas the FSU-family parameterizations produce comparatively smoother variations with $\sqrt{s_{\rm NN}}$. The TM1 and TM1e parameterizations yield intermediate behavior.

{We have further quantified the agreement between the theoretical predictions and the available BES/STAR measurements by evaluating the average deviation $\Delta_i$ over the experimentally accessible beam-energy range. The FSU-family parameterizations exhibit relatively small deviations from the measured heavy-ion observables i.e. cumulant ratios $i = \chi_3/\chi_1,\ \chi_3/\chi_2,\ {\rm and}\ \chi_4/\chi_2$. However, their consistency with the heavy-ion observables alone does not guarantee simultaneous compatibility with the astrophysical constraints. When the constraints from the BES/STAR measurements are considered together with nuclear-matter saturation properties and neutron-star observations, the FSU2R parameterization provides a consistent description within the set of models investigated here. In this sense, the combined analysis illustrates the importance of confronting heavy-ion fluctuation observables with independent constraints from finite nuclei and compact-star observations rather than relying on a single class of observables.}

The present analysis is restricted to nucleonic matter and therefore provides a well-defined baseline for investigating the effects of additional degrees of freedom. 
A natural extension is to incorporate hyperonic degrees of freedom. The appearance of hyperons soften the EOS and alter the density dependence of the mean fields, potentially producing characteristic modifications in higher-order fluctuations and hence cumulant ratios. 

\section*{Data availability}
No new data were generated or analyzed in the present study.

\bibliographystyle{unsrt}
\bibliography{zzbib}
\end{document}